\documentclass[aps,nofootinbib,twocolumn,superscriptaddress]{revtex4}
\usepackage{amsfonts,amssymb,amsmath}            
\usepackage[]{graphics,graphicx,epsfig}            
\usepackage{amsthm}
\def\identity{\leavevmode\hbox{\small1\kern-3.8pt\normalsize1}}

\newtheorem{lemma}{Lemma}
\newcommand{\ket}[1]{\left | #1 \right\rangle}
\newcommand{\bra}[1]{\left \langle #1 \right |}
\newcommand{\half}{\mbox{$\textstyle \frac{1}{2}$}}

\newcommand{\proj}[1]{\ket{#1}\bra{#1}}
\renewcommand{\epsilon}{\varepsilon}
\begin{document}

\title{Interfacing with Hamiltonian Dynamics}
\date{\today}

\author{Alastair \surname{Kay}}
\affiliation{Max-Planck-Institut f\"ur Quantenoptik, Hans-Kopfermann-Str.\ 1,
D-85748 Garching, Germany}
\affiliation{Centre for Quantum Computation,
             DAMTP,
             Centre for Mathematical Sciences,
             University of Cambridge,
             Wilberforce Road,
             Cambridge CB3 0WA, UK}
\affiliation{Centre for Quantum Technologies, National University of Singapore, 3 Science Drive 2, Singapore 117543}

\begin{abstract}
Simple constructions and protocols are demonstrated to allow the implementation of universal quantum computation on an arbitrarily large quantum system by controlling a fixed number of spins, vastly reducing the engineering requirements in comparison to a direct implementation of the circuit model. Analytic pulse sequences are derived for deterministic gate operation and system cooling which only require pulse strengths that are comparable to the intrinsic Hamiltonian of the system.
\end{abstract}

\maketitle

\section{Introduction}

The circuit model for quantum computation has been instrumental in the development of a plethora of experimental proposals for the implementation of quantum computation. One merely has to demonstrate how to implement an arbitrary single-qubit rotation on any qubit, and an entangling two-qubit gate between any pair of qubits. However, the circuit model should not be considered a natural paradigm for the control of arbitrary quantum systems, merely a useful tool for proving the universality of the system in question. Indeed, it is often impossible, or extremely undesirable, to implement such a scheme directly. For example, single spin manipulation is extremely challenging in systems such as optical lattices. In other schemes, extremely high fidelity single-qubit operations are possible, but the number of control elements required for the individual addressing of each and every qubit in a register is prohibitive.

This prompts the question of how little control is required, which has been answered by proving that Hamiltonian evolution, even for translationally invariant and rotationally invariant Hamiltonians, can implement arbitrary quantum computations \cite{karl,Kay:08,followup}. However, these constructions require spins which are larger than qubits, or interactions with some spatial extent beyond nearest neighbor. Moreover, one must be able to impose suitable initial conditions (a product state), which could in themselves require the local control that may not be available. Finally, error correction and fault-tolerance have not been built into these schemes, although there are some closely related schemes based around quantum cellular automata \cite{shepherd} which can be shown to be fault-tolerant \cite{Kay:thesis}. Reintroducing a small amount of control into such Hamiltonian systems can markedly reduce the overheads involved in the constructions. One such example is global control, where several different Hamiltonian evolutions can be pulsed, and their sequencing is the added control that allows the reduction to nearest-neighbor qubit systems in a fault-tolerant setting (see \cite{Lloyd:1993a,Kay:thesis} and references therein).

In this paper, we will study another possible reduction, where control of a fixed number of spins is retained, and show how, through manipulation of these, quantum computation can be achieved in the entire quantum system. This is a plausible control paradigm for early experiments in quantum computation. Such a scheme was referred to as a Universal Quantum Interface (UQI) by Lloyd and co-authors \cite{UQI}. These and others have subsequently proved that almost all quantum systems can be controlled in this way, including nearest-neighbor translationally invariant qubit systems \cite{Burgarth:07}. However, once the proof of existence is completed, these schemes fall back on numerical techniques for determining the control sequence for a system of specific size, and typically reveal little about the solution for systems of different size, or the efficiency of determining these sequences. Instead, we concentrate on slightly more complex systems where we are able to give concrete protocols for arbitrary system size. We are also able to discuss how to handle experimental shortcomings such as finite implementation times for the control sequences. This presents a number of improvements over \cite{Kay:08}, where the construction technique was first introduced as a tool for designing more complex Hamiltonians. For example, our most basic scheme still involves a nearest-neighbor translationally invariant Hamiltonian acting on 4-dimensional spins, and control is over 3 of the spins. However, we now only need single-spin control of these spins, rather than needing to make them interact. We also discuss how these chains can be deterministically cooled from their initial, unknown, state, to the state required for the computation.

\section{Quantum State Transfer}

\subsection{Basics}

We start our study by reviewing the basic components that we will use in Sec.~\ref{sec:main} to formulate the UQI systems. We shall quote several results, with little justification. Their relevance will become apparent as we start to construct UQI schemes. The principal element here is the study of quantum state transfer. This was initiated by Bose \cite{Bos03}, who took a simple nearest-neighbor Hamiltonian on a 1D chain of $N$ qubits
$$
H_{ST}=\half(Z_1+Z_N)+\half\sum_{i=1}^{N-1}X_iX_{i+1}+Y_iY_{i+1},
$$
assumed the chain was initially prepared in the state $\ket{\psi}\ket{0}^{\otimes N-1}$, and demanded how well the unknown single-qubit state $\ket{\psi}$ can be transferred to the other end of the chain. This was facilitated by the observation that
$$
\left[H_{ST},\sum_{i=1}^NZ_i\right]=0,
$$
so the Hilbert space of the Hamiltonian decomposes into subspaces of fixed excitation number (number of $\ket{1}$s)\footnote{Note that $H_{ST}$ is not the Hamiltonian considered by Bose. However, in the single excitation subspace, the two are identical.}. The state $\ket{0}^{\otimes N}$ is an eigenstate, and thus the task is to transfer from $\ket{1}\ket{0}^{\otimes N-1}$ to $\ket{0}^{\otimes N-1}\ket{1}$. In fact, for the purposes of designing UQIs, it is possible to entirely neglect the transfer of the $\ket{0}$ component of $\ket{\psi}$, and just consider the transfer of $\ket{1}$. Developing the theory of state transfer will subsequently allow a mapping from the UQI Hamiltonians into a state transfer system $H_{ST}$ such that $\ket{1}\ket{0}^{\otimes N-1}$ can be considered the start of the computation, and $\ket{0}^{\otimes N-1}\ket{1}$ is the end of the computation. Such a transformation can be recognized by the way the Hamiltonian acts on basis states,
$$
H_{ST}\ket{n}=\ket{n-1}+\ket{n+1},
$$
where, in this case, the basis states are defined to be $\ket{n}=\ket{0}^{\otimes n-1}\ket{1}\ket{0}^{\otimes N-n}$.
In this situation, one can prove several basic facts, including
\begin{lemma}
State transfer can be achieved with a probability $O(N^{-1/3})$ in a time $O(N)$ \cite{Bos03}. \label{lemma:1}
\end{lemma}
\begin{lemma}
The arrival of $\ket{\psi}=\ket{1}$ can be detected by measuring spin $N$. A strategy of measure and repeat until the state has arrived gives a failure probability of less than $\varepsilon$ in a time $O(N^{5/3}\log(\varepsilon))$ \cite{Bos04}. \label{lemma:2}
\end{lemma}
\begin{lemma}
The Hamiltonian $H_{ST}-Z_1-Z_N$ can be mapped to a system of free (non-interacting) fermions.
\end{lemma}
Given that the fermions do not interact, one can readily translate results on state transfer from the single excitation subspace to multiple excitation subspaces, provided the phase properties of fermion exchange are correctly handled \cite{Christandl:2004a}. A useful tool for this is the wedge product \cite{Osborne,Kay:2005e}. For example $\ket{1}\wedge\ket{2}$ denotes the presence of excitations on spins 1 and 2. The critical property of this product is that $\ket{1}\wedge\ket{1}=0$.

\subsection{Wavepackets}

There are a number of strategies that one could utilize to improve the arrival of a single excitation. Haselgrove \cite{haselgrove04} gave a numerical technique, applicable to very general networks, for finding how Alice should distribute the excitation across the first $N_A$ spins such that it is optimally transmitted to Bob, who controls the last $N_B$ spins of the chain, the idea being to work in the regime where $N_A+N_B\ll N$. However, consideration of the special case of a linear chain, enables the derivation of rigorous lower bounds, utilizing the following fact:
\begin{lemma}
In an infinite chain, away from the ends of the chain, the optimal encoding of a single excitation is in a Gaussian wavepacket of spread $\Delta\sim N^{-1/3}$, where $N$ is the distance of transmission \cite{osborne04}.
\end{lemma}
The wavepacket that has the maximum group velocity for the system, and is minimally dispersive, can be written using the previously defined  position basis as
\begin{equation}
\ket{\psi_G}=\frac{1}{\sqrt{\Delta\sqrt{\pi}}}\sum_{n=-\infty}^{\infty}e^{-(n-x_0)^2/(2\Delta^2)-i\pi n/2}\ket{n}, \label{eqn:gaussian}
\end{equation}
which is centered on spin $x_0$ at time $t=0$, and its center is located around $x_0+2t$ at any later time $t$. The speed is dimensionless due to the choice of units of, $\hbar=1$, and the suppression of the energy scale of $H_{ST}$.

In addition to the deterministic arrival and low dispersion of the wavepacket, a further advantage is that the wavepacket is well localized. This means that multiple excitations can be sent at a rate of $O(N/\Delta)$, without degradation of the signal \cite{Osborne,Kay:2005e}.

\subsection{Time Control and Wavepacket Injection}

\begin{figure*}
\begin{center}
\includegraphics[width=0.8\textwidth]{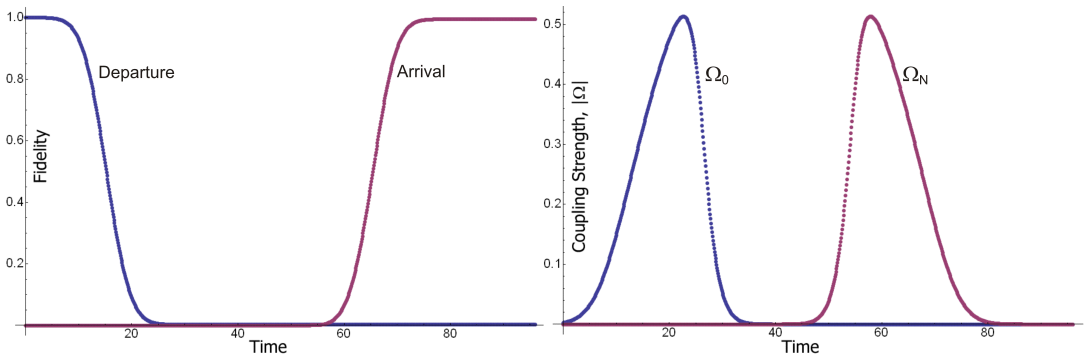}
\end{center}
\vspace{-0.5cm}
\caption{Plot, as a function of time, for (a) Fidelity of arrival and departure of states $\ket{2}^{\otimes 100}$ and $\ket{0}^{\otimes 100}$ respectively for $f=3$ and $\Delta=10$, giving an arrival fidelity of $99.45\%$, and (b) coupling strengths $\Omega_0$ and $\Omega_N$. The final fidelity can be increased by increasing $\Delta$.} \label{fig:pulses}
\vspace{-0.5cm}
\end{figure*}

While the optimal wavepacket for transmission through the central region of a long chain is known, this is of little benefit since we must find the encoding at the end of the chains, as well as needing to find how to inject the wavepacket onto the chain, and remove it at the end. To this end, consider a second situation introduced in \cite{haselgrove04}, where time-varying control at the ends of the chains is available,
\begin{eqnarray}
2H_{TV}&=&\Omega_0(t)(X_0Y_1+X_0Y_1)+\Omega_N(t)(X_NY_{N+1}+X_NY_{N+1})    \nonumber\\
&&+\sum_{i=1}^{N-1}X_iX_{i+1}+Y_iY_{i+1}.  \nonumber
\end{eqnarray}
This time-varying control can be used to simulate an infinite spin chain i.e.~to make the $N$ spins of the chain behave as if they were in the middle of a much longer chain, introduce a Gaussian wavepacket on one end, and remove it from the other end at a time $O(N)$ later. The state of the chain to be simulated, as a function of time, $t$, is
$$
\sum_{n=-\infty}^{\infty}\psi_n(t)\ket{n}
$$
Similarly, the state of the actual chain is
$$
\sum_{n=0}^{N+1}\phi_n(t)\ket{n},
$$
where the aim is to ensure that $\psi_n(t)=\phi_n(t)$ for all $t\geq0$ and $n=1\ldots N$. At $t=0$, the packet starts in a negative position, $x_0=-f\Delta$, where $f$ encapsulates an error parameter determining how many standard deviations of the wavepacket we wish to use, ultimately influencing the error of the simulation. In the time $t=f\Delta+N/2$ the wavepacket moves along the infinite chain with a ballistic motion, ending up at a position $N+f\Delta$, outside the region of interest. Using Eqn.~(\ref{eqn:gaussian}),
$$
\psi_n(t)=\frac{1}{\sqrt{\Delta\sqrt{\pi}}}e^{-(n-x_0-2t)^2/(2\Delta^2)-i\pi n/2}.
$$
In order for the actual chain to simulate the infinite chain, it suffices to ensure that $\psi_n(0)=\phi_n(0)$ and $\frac{d\psi_n}{dt}=\frac{d\phi_n}{dt}$ for all times. This is convenient because the Hamiltonians are identical in the two cases, except at the ends of the chain. Thus, we have
$$
i\frac{d\phi_1}{dt}=\Omega_0\phi_0+\phi_2=\psi_0+\psi_2,
$$
from which
$$
\Omega_0=\frac{\psi_0}{\phi_0}
$$
is readily derived.
It is worth noting that $|\phi_0|^2=\sum_{n=-\infty}^0|\psi_n|^2,$ which could be used to determine $|\Omega_0|$. However, an alternative derivation\footnote{In the limit of $\Delta\rightarrow\infty$, the two derivations give the same pulse. In fact, for small $\Delta$, this derivation, by setting $|\phi_0|^2=\int_{n=-\infty}^{\half}|\psi(x)|^2dx,$, seems to give better results numerically, but the pulses extend over longer time} (since the system is not continuous), which allows the modulus sign to be removed, derives from
$$
i\frac{\partial \phi_0}{\partial t}=\Omega_0(t)\psi_1(t) \Rightarrow \half i\frac{\partial \phi_0^2}{\partial t}=\psi_0(t)\psi_1(t),
$$
which ultimately reveals that
$$
\Omega_0(t)=\frac{e^{-(x_0+2t)^2/2\Delta^2}}{\sqrt{\Delta\sqrt{\pi}}\sqrt{1-\half e^{-1/4\Delta ^2}\left(1+\text{Erf}\left(\frac{2t+x_0-\half}{\Delta}\right)\right)}}.
$$
where $\text{Erf}(z)$ is the error function. At $t=0$, $\Omega_0(t)\approx 0$, and for all times greater than $O(\Delta)$ (i.e. once the bulk of the wavepacket is on the chain), this is true again. In the meantime, $\Omega_0(t)$ is a smooth function that is never greater than 1, as can be seen from the numerical example depicted in Fig.~\ref{fig:pulses}. The total time for the sequence is $O(N+N^{1/3}\log\varepsilon)$, where the second term incorporates the influence of increasing the accuracy by increasing $f$.

Similarly, one can derive that
$$
\Omega_N(t)=\frac{e^{-(x_0+2t-N-1)^2/2\Delta^2}}{\sqrt{\Delta\sqrt{\pi}}\sqrt{1-\half e^{-1/4\Delta ^2}\left(1-\text{Erf}\left(\frac{2t+x_0-N-\half}{\Delta}\right)\right)}}.
$$
The boundary condition that is imposed on $\phi_{N+1}(t)$ is $\phi_{N+1}(\infty)=(-i)^{N+1}$, which is due to a phase issue that occurs during a state transfer protocol (see, for example, \cite{Kay:2004c}).

Although concentrating on the specific task of transferring an excitation, this analytic solution is applicable to the transfer of an unknown single qubit state $\ket{\psi}$ initially stored on spin 0 onto spin $N+1$ simply due to linearity -- the $\ket{1}$ state transfers, and the state containing no excitations is an eigenstate. In the remainder of this subsection, we will show that the protocol is very robust against control errors and, unlike other state transfer schemes, the available control allows the preparation of the initial state of the system rather than having to assume an additional cooling mechanism.

\subsubsection{Robustness} \label{sec:robust}

\begin{figure*}
\begin{center}
\includegraphics[width=0.8\textwidth]{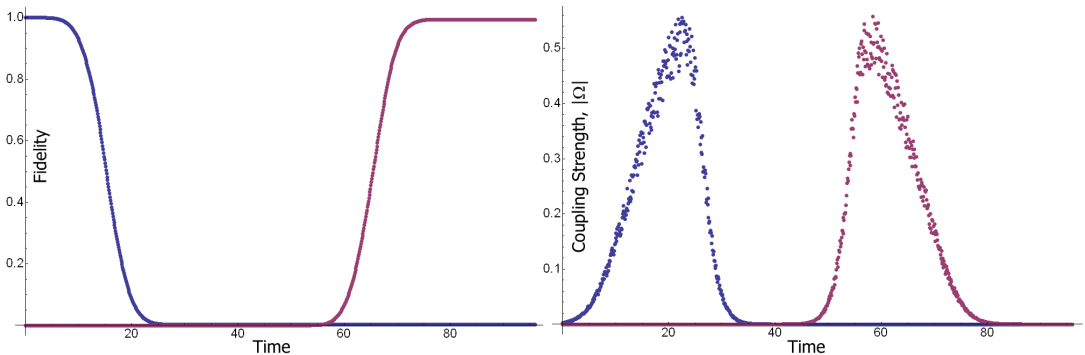}
\end{center}
\vspace{-0.5cm}
\caption{Plot, as a function of time, for (a) Fidelity of arrival and departure of states $\ket{2}^{\otimes 100}$ and $\ket{0}^{\otimes 100}$ respectively for $f=3$ and $\Delta=10$, giving arrival fidelity of $99.37\%$, and (b) coupling strengths $\tilde\Omega_0$ and $\tilde\Omega_N$, which are randomly varied from their ideal forms (Fig.~\ref{fig:pulses}) within the range $\pm 10\%$. To be compared with Fig.~\ref{fig:pulses}.} \label{fig:randoms}
\vspace{-0.5cm}
\end{figure*}

The described protocol usefully indicates how a number of very relevant physical effects can be incorporated into the scheme, such as finite laser power. There are two further experimental concerns, decoherence and imperfect control operations. We shall start by numerically testing the effects of imperfect control operations. Take, as an example, the case of $N=100$, $\Delta=30$ and $f=3$. To simulate the time evolution, we split the sequence into time steps of 0.1 (compared to the speed of the wavepacket, which is 2), and assume that the values of $\Omega_{0,N}(t)$ are constant over that timescale. For the ideal pulse sequence, the success probability is $99.94\%$. This probability tends to $100\%$ as $f\rightarrow\infty$ and $\Delta/N^{1/3}\rightarrow\infty$. Even after failure, all is not lost because we can continue to perform a heralded state transfer protocol until successful.

\begin{figure}[!b]
\begin{center}
\includegraphics[width=0.4\textwidth]{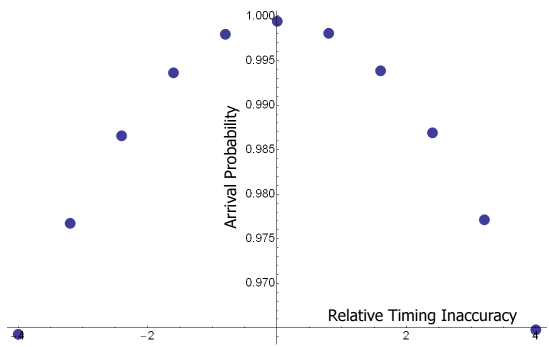}
\end{center}
\vspace{-0.5cm}
\caption{Arrival fidelity as a function of timing error, $t_N-t_0$, for a chain of 100 spins, $\Delta=30$, $f=3$.} \label{fig:timing}
\vspace{-0.5cm}
\end{figure}

First, we consider timing errors i.e.~$\tilde \Omega_i(t)=\Omega_i(t-t_i)$. Evidently, $t_0$ determines when the protocol starts, and so the only relevant parameter is the relative time $t_N-t_0$. A plot of calculated values is given in Fig.~\ref{fig:timing}.

The next imperfection that we consider is the strength of the laser pulse. For example, with a systematic offset of $5\%$, i.e.~$\tilde\Omega_i(t)=1.05\Omega_i(t)$, the arrival probability is still $99.2\%$. Finally, we consider fluctuations of the control fields, by a random amount at each time step, where we considered two different cases; where the maximum fluctuation was up to $10\%$ of the ideal field at that time (arrival probability $0.9988(1)$), and by up to $0.01$ (compared to the maximum field strength of approximately 0.3), independent of the value of the intended field at that time (arrival probability $0.9986(2)$). An example of one instance of this evolution is given in Fig.~\ref{fig:randoms}. These results suggest a large degree of tolerance to control errors.

\subsubsection{System Initialization} \label{sec:cooling}

In addition to using the temporal control for injecting wavepackets into the system, it can be utilized to provide an effective cooling from an unknown initial state to the state $\ket{0}^{\otimes N}$. We shall assume that we have the ability to measure spins 0 and $N+1$, and to reset them to the $\ket{0}$ state. As before, the time-varying controls at the end of the chain can be used to simulate a larger chain, designed such that, with high probability, after a fixed period of time, excitations leave the part of the chain that they're initially localized on. We start by designing a protocol under the assumption that there is a single excitation somewhere on the chain. Working sequentially, starting with spin 1, pulse sequences can be designed so that if an excitation is there, it moves off onto spin 0 or spin $N+1$, and can be reset. When spin 0 is measured, it reveals whether the excitation was removed or not. If not, the protocol continues. If the excitation had originally been on spin 2, it is no longer localized there. However, the state that it would be in after the first pulse sequence can be found via an efficient classical computation. So, a pulse sequence can be designed to remove it. This continues for all $N$ spins, after which the single excitation is guaranteed to have been removed.

In the case where multiple excitations are present, exactly the same protocol serves to remove all excitations, except that we do not stop the protocol once a single excitation has been removed. For example, let's assume that there were originally 3 excitations, one on each of spins 1, 2 and 4, which is denoted by $\ket{1}\wedge\ket{2}\wedge\ket{4}$. The first 4 pulse sequences that we go through can be described by $U_1$ to $U_4$ respectively. After the first evolution,  $U_1\ket{1}=\ket{0}$, which is detected and removed, leaving $U_1\ket{2}\wedge U_1\ket{4}$ (essential to this is the fact that $\bra{0}U_1\ket{2}=\bra{0}U_1\ket{4}=0$). After the second evolution, we have
$$
U_2U_1\ket{2}\wedge U_2U_1\ket{4}=\ket{0}\wedge U_2U_1\ket{4},
$$
Again, since $\bra{0}U_2U_1\ket{4}=0$, this is just equal to $U_2U_1\ket{4}$. After the third pulse is used, no excitation is detected on spin 0, but on the fourth we have $U_4U_3U_2U_1\ket{4}=\ket{0}$, and all the excitations are successfully removed.

All that remains is to specify the chain to simulate with $\Omega_i(t)$. One strategy is to use the perfect mirroring chain introduced in \cite{christandl}, where it was proved that under the action of the Hamiltonian
$$
H_{PST}=\half\sum_{n=1}^{M-1}\sqrt{n(M-n)}(X_nX_{n+1}+Y_nY_{n+1})
$$
for a time $\pi/2$, an excitation initially on spin $n$ arrives perfectly on spin $M+1-n$. For a chain with a large number of qubits $M$, the central $N$ spins are approximately uniformly coupled, with a value of $M/2$. Thus, a good approximation for simulation of this system can be achieved, after rescaling to $2H_{PST}/M$. By selecting the mirroring point to be just offset from the center of the chain, excitations get moved off the ends. The exact solutions for the functions $\psi_m(t)=\bra{m}e^{-iH_{PST}t}\ket{n}$ can be found in \cite{Christandl:2004a} and, as such, we are able to take the limit of $M\rightarrow\infty$ to achieve an arbitrarily accurate simulation. To prove this, consider the perturbation
$$
\delta H=\!\!\!\!\!\!\sum_{n=(M-N)/2}^{(M+N)/2}\!\!\!\!\!\!(M/2-\sqrt{n(M-n)})(\ket{n}\bra{n+1}+\ket{n+1}\bra{n})
$$
acting on $H_{PST}$. By ensuring that $\delta H$ retains the mirror symmetry of the original Hamiltonian, the analysis is simplified because it is only the eigenvalues that affect the fidelity of state transfer \cite{Kay:2004c}. Provided the matrix elements of $\delta H$ are much smaller than 1, i.e.~$M\gg N^2$, perturbation theory allows us to calculate how the eigenvalues are shifted, using $\bra{\lambda_n}\delta H\ket{\lambda_n}$, where $\ket{\lambda_n}$ are the unperturbed eigenvectors. The error can be bounded by using the eigenvector which has the largest elements in the middle of the chain,
$$
\ket{\lambda_0}=\frac{1}{2^{(M-1)/2}}\sum_{n=1}^M\sqrt{\binom{M-1}{n-1}}\ket{n}.
$$
Taking the large $N$ limit and using Stirling's approximation, we find that
$$
\bra{\lambda_0}\delta H\ket{\lambda_0}\sim\frac{N^3}{M^{5/2}}.
$$
Provided this is small, all higher order terms from perturbation theory are even smaller and can be neglected. Small errors of size $\delta$ in the eigenvalues contribute to a reduction in arrival fidelity of the state mirroring of $O(\delta^2)$ \cite{Kay:2005e}, and thus we are guaranteed that in the large $M$ limit the error tends to 0.

\subsection{A Second Time-Controlled Scenario} \label{sec:spin2}

\begin{figure}[!b]
\begin{center}
\includegraphics[width=0.3\textwidth]{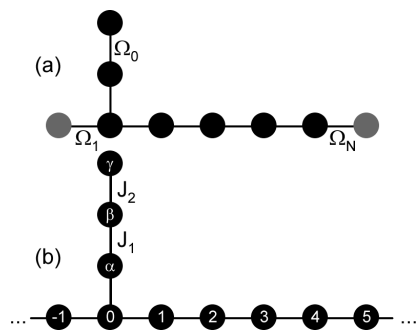}
\end{center}
\vspace{-0.5cm}
\caption{(a) Time-controlled spin chain, used to simulate the fixed spin chain depicted in (b), where $\ldots$ indicates the presence of a chain of $N_A$ (left) or $N_B$ (right) qubits, either of which may be taken to $\infty$.} \label{fig:new}
\vspace{-0.5cm}
\end{figure}

There is a second instance that we will be interested in, where the scheme is no longer a perfect chain, but the network depicted in Fig.~\ref{fig:new}(a). Again, the aim is to achieve perfect excitation transfer between the shaded spins, or to place a lower bound on how well this can be achieved. We would aim to do this, as before, by simulating a much longer chain. However, when trying to transmit a Gaussian wavepacket through the system in Fig.~\ref{fig:new}(b), part of the packet gets reflected at the T-junction, and transmission is no longer perfect. A heuristic explanation for how to get around this problem is that we know that wavepackets can be created that travel along the chain with speed 2. When they hit the junction, they are partially reflected. If the side that Alice controls is short (i.~e.~$N_A\nrightarrow\infty$), and the side that Bob controls long, then the first reflection will rebound off Alice's end, and impinge once again on the T-junction, part of which will get transmitted. Everything that gets transmitted can be received by Bob, provided his region of the chain is long enough that the wavepacket does not reflect and move out again. If the transmission probability is $T$, then after $m$ such cycles (requiring a time $mN_A$) some proportion which is roughly $\sum_{i=0}^{m-1}T(1-T)^i$ has passed beyond the imperfection, and can be gathered by Bob after a further time $N/2$, provided $N_B>2(m-1)N_A$. For a probability of failure $\varepsilon$, we would only require $m\sim\log\varepsilon$ provided $T$ is a constant\footnote{After calculating the variation of $T$ as a function of the energy, this can be derived more rigorously, taking into account the fact that the reflected wavepacket is no longer Gaussian, and therefore its transmission probability is different.}.

The first step towards proving this is to show that for a half-infinite chain, when a wavepacket is incident on the end, it is reflected as a wavepacket that continues to evolve as if the boundary was not present. To see this, consider an infinite chain where the initial state is an equal superposition of two wavepackets (taken, for simplicity of exposition, to be Gaussian) positioned at $x_0$ and $-x_0$, traveling towards each other with equal speeds $v$,
$$
(\ket{\psi_G(x_0,-v)}-\ket{\psi_G(-x_0,v)})/\sqrt{2}.
$$
By linearity, it is clear how these packets evolve; they just move smoothly past each other and, at long times, are two Gaussian wavepackets moving away from each other at equal speeds,
$$
(\ket{\psi_G(x_0-vt,-v)}-\ket{\psi_G(-x_0+vt,v)})/\sqrt{2}.
$$
The Hamiltonian of the infinite system in the single excitation subspace commutes with the symmetry operator
$$
S=\sum_{n=1}^{\infty}\ket{n}\bra{-n}+\proj{0}
$$
and, as such, the Hilbert space can be decomposed into two subspaces, symmetric and antisymmetric. The described wavepacket is antisymmetric, and thus its evolution is described entirely within that subspace, which is precisely that of a Gaussian wavepacket on a half-infinite chain impinging on the boundary. Thus, we learn that wavepackets reflect perfectly\footnote{By which we mean that when the wavepacket is a long distance from the boundary, it is Gaussian}, although they acquire a phase of $\pi$. It still remains to prove the expected reflection off the T-junction. Having proved what happens at the boundaries, it is now sufficient to take the size of the simulated systems $N_A,N_B\rightarrow\infty$, and show what happens when a wavepacket is incident on the imperfection. This proof will closely follow the work of Childs, \cite{childs}. The first step is to find the eigenvalues and eigenvectors for the network in Fig.~\ref{fig:new}(b). Ignoring the T-junction, the unnormalized eigenvectors would take the form
$$
\sum_{x=-\infty}^{\infty}e^{-ikx}\ket{x}
$$
with energies $2\cos k$. However, the T-junction causes partial reflection of the wave. Thus, we write the new eigenvectors incorporating the amplitudes for transmission and reflection,
$$
\sum_{x=-\infty}^{-1}(e^{-ikx}+R_ke^{ikx})\ket{x}+\sum_{x=1}^{\infty}T_ke^{-ikx}\ket{x}+\sum_{i\in\{0,\alpha,\beta,\gamma\}}a_i\ket{i},
$$
and solve the set of linear equations such that the energy is still $2\cos k$. This reveals that the transmission probability for a wave with momentum $k$ is
$$
|T_k|^2=\frac{4\sin^2(2k)(J_1^2+J_2^2-4\cos^2k)^2}{4\sin^2(2k)(J_1^2+J_2^2-4\cos^2k)^2+(J_2^2-4\cos^2k)^2}.
$$
By selecting $J_2^2=4\cos^2k$, perfect transmission can be realized for a specific value of $k$. To use the minimally dispersive wavepacket, with maximum transmission rate, $k=\pi/2$ is the ideal choice, so $J_2=0$ is selected (i.e.~$\Omega_0=J_1$ is constant). Expanding for small $\delta k$ about this point, we find that
$$
|T_{\pi/2+\delta k}|^2=1-\frac{4}{J_1^4}\delta k^2+O(\delta k^3).
$$
For a Gaussian wavepacket of width $\Delta$, the width in $k$-space is roughly $1/\Delta$, and thus the transmission probability $\sim 1-O(N^{-2/3})$, so in the limit of large $N$, the success probability tends to 1. In the instance where only the right-hand side of Fig.~\ref{fig:new}(b) is semi-infinite, and the left-hand side only extends to $-N_A$, any component that is reflected off the T-junction is perfectly reflected off the finite end, and is subsequently incident on the T-junction again. By waiting long enough to collect $m$ such repetitions, the success probability is enhanced to
$$
1-\frac{4}{J_1^4\Delta^{m+1}(m+1)!}+O\left(\frac{1}{\Delta^{m+3}}\right).
$$
The choice of $m$ is related to the length of time spent collecting the wavepacket, $m\sim (t/2-N)/(2N_A)$. Thus, the total scaling is also $O(N+N^{1/3}\log\varepsilon)$ for a maximum error of $\varepsilon$, and the entire wavepacket still has a width $O(N^{1/3})$, so the scaling of the rate of transmission is the same as the previous scheme.

\section{Designing a UQI} \label{sec:main}

Having stated and derived all the basic tools that are required for the construction of a UQI within the framework of state transfer, we are in a position to start designing Hamiltonians whose dynamics can lead to quantum computation when aided by dynamical control over a fixed number of spins, by proving mappings to the state transfer results. In order to prove universal quantum computation, it is sufficient to show how to perform any single-qubit rotation on any qubit, and how to perform just one entangling gate between an arbitrary pair of qubits. Given that we retain control over at least two qubits, in principle it suffices to show how to place any pair of qubits on those two control qubits, and the arbitrary control over them allows the implementation of the desired gates. So, all we need is a simple permutation operation. This motivated the original UQI construction in \cite{Kay:08}, acting on a chain of $N$ 4-dimensional spins, whose local Hilbert space is decomposed into two qubits, denoted $a$ and $b$,
$$
H_1=\half\sum_{n=1}^{N-1}(XX+YY)_{n,n+1}^a\otimes\text{SWAP}_{n,n+1}^b.
$$
To see how this works, start by observing that
$$
\left[H_1,\sum_{n=1}^NZ^a_n\right]=0,
$$
and adopt a basis in the single excitation subspace of the $a$ qubits of $\ket{n}$, $n=1\ldots N$, denoting the presence of the single excitation on spin $n$. The action of $H_1$ on these basis states can be written as
$$
H_1\ket{n}^a\ket{\psi_n}^b=\ket{n-1}^a\ket{\psi_{n-1}}^b+\ket{n+1}^a\ket{\psi_{n+1}}^b,
$$
where $\ket{\psi_{n+1}}=\text{SWAP}_{n,n+1}\ket{\psi_n}$, and thus this Hamiltonian maps into the state transfer system $H_{ST}-Z_1-Z_N$. With control over spins 1 and $N$, $Z_1$ and $Z_N$ can be incorporated. Thus, by invoking Lemma \ref{lemma:1}, if the system were to start in $\ket{1}^a\ket{\psi}^b$, then after a time $O(N)$, with probability $O(N^{-1/3})$, it would be in state $\ket{N}^a\left(\text{SWAP}_{N-1,N}\ldots\text{SWAP}_{1,2}\ket{\psi}\right)^b$, which is a cyclic permutation of the $b$ qubits. Multiple repetitions of this protocol enable any arbitrary qubit to be placed on spin 1. Moreover, by measuring to see if an excitation is on $\ket{N}^a$, the arrival is heralded, and the protocol of Lemma \ref{lemma:2} gives arrival with failure probability less than $\varepsilon$ in a time $O(N^{5/3}\log(\varepsilon))$.

Similarly, we could start an excitation on $\ket{2}^a\ket{\psi}^b$ and transfer it to $\ket{N}^a\left(\text{SWAP}_{N-1,N}\ldots\text{SWAP}_{2,3}\ket{\psi}\right)^b$, which also implements a cycling operation, but only on qubits 2 to $N$. Thus, any pair of qubits can be placed on the first two spins by controlling only the first 2 spins and spin $N$. In order to implement a one or two-qubit gate requires a time $O(N^{8/3}\log(\varepsilon))$.

There are now a number of potential improvements that we would like to introduce. Firstly, we might like to restrict our control of spins to being strictly local, i.e. no two-qubit interaction. We do this by recalling a result of \cite{twamley}, whereby $N+1$ repetitions of global applications of controlled-phase ($CP$) gates and Hadamard gates ($H$) gives the mirroring operation between the $N$ qubits (i.e.~the state of spin $n$ is mapped to that of $N+1-n$). By altering $H_1$, replacing SWAP with the new operation $U=(H\otimes\identity)\cdot CP$, we get
$$
H_2=\sum_{n=1}^{N-1}\ket{01}\bra{10}^a_{n,n+1}\otimes U +h.c.
$$
Note that because $U\neq U^\dagger$, it has become necessary to split the $\half(XX+YY)$ in order to maintain the Hermitian nature of the Hamiltonian. We interpret this as being that when the excitation steps to the right, it implements $U$ on the $b$ register, and when it steps to the left, it implements $U^{\dagger}$, which means that the position of the excitation alone is sufficient to reveal what sequence of gates has been implemented. Now when we perform a heralded state transfer protocol from 1 to $N$, the implemented unitary is ${\tilde U}=(H_1\otimes\ldots\otimes H_{N-1})(CP_{N-1,N}\ldots CP_{1,2})$, which is the aforementioned global application of controlled-phases followed by Hadamards, except that a Hadamard needs to be applied on the $b$ qubit of spin $N$. Thus, $N+1$ repetitions of the protocol implements the mirroring operation on spins 1 to $N$. Similarly, performing transfer from 2 to $N$ implements mirroring on spins 2 to $N$. The combination of the two is the previous cyclic permutation; thus we recover the ability to place any single qubit on spin 1, although it now means that a one-qubit gate requires time $O(N^{11/3}\log(\varepsilon))$. To achieve a two-qubit gate, we run the state transfer protocol from 1 to $N$, apply a Hadamard gate on qubit $1^b$, and then run the state transfer protocol from $N$ to 1, which implements the inverse unitary. Thus, ${\tilde U}^{\dagger}H{\tilde U}$ is a two-qubit gate acting on spins 1 and 2 which is locally equivalent to controlled-NOT. So, we have an entangling gate between spins 1 and 2, and cycle. However, the controlled-NOT can be composed to give SWAP, so that as the qubits are cycled, SWAP can cause a single qubit to stay on spin 1 while the others cycle. Therefore, this can be used to implement the two-qubit gate between any pair of qubits.

\subsection{Realistic Pulses}

So far, we have assumed that we can place excitations on spin 1, remove them from spin $N$ etc., arbitrarily quickly, at least compared to the inverse of the Hamiltonian interaction strength. This is not a very realistic condition, and one that can be remedied. One of the problems with a slow laser pulse which converts $\ket{0}$ to $\ket{1}$ on a particular spin is that when it has been partially rotated, the Hamiltonian is already moving that part of the excitation onto other spins. As a consequence, it is possible that the pulse would introduce multiple excitations. One trick to avoid this involves introducing a third level, $\ket{2}$, to system $a$ such that when an excitation hops to the right due to the Hamiltonian, it leaves behind it a $\ket{2}$ rather than a $\ket{0}$. Thus, a laser $\Omega_0(\ket{0}\bra{1}+\ket{1}\bra{0})$ cannot create excitations on the $\ket{2}$ component. Such a Hamiltonian takes the form
$$
H_3=\sum_{n=1}^{N-1}\ket{21}\bra{10}^a_{n,n+1}\otimes U+\ket{01}\bra{12}^a_{n,n+1}\otimes \identity +h.c.
$$
When enhanced by the laser fields
$$
H_L=\Omega_0(\ket{0}\bra{1}+\ket{1}\bra{0})^a_1+\Omega_N(\ket{2}\bra{1}+\ket{1}\bra{2})^a_N,
$$
we can now prove the mapping to the state transfer scheme of $H_{TV}$, so $\ket{0}^{\otimes N}$ is coupled to $\ket{1}\ket{0}^{\otimes N-1}$ with strength $\Omega_0(t)$, and the central states are of the form $\ket{2}^{\otimes n-1}\ket{1}\ket{0}^{\otimes N-n}$. Finally, the $\Omega_N$ pulse leaves the system in the state $\ket{2}^{\otimes N}$, from which it must be reset before the next $\tilde U$ can be applied. This is achieved by applying a laser on spin 1 to convert the $\ket{2}$ into a $\ket{1}$, and on spin $N$ to convert a $\ket{1}$ into a $\ket{0}$. The second term in $H_3$ then takes care of the rest.

For the part of the sequence where we start the excitation on spin 2, the scheme maps into that of Sec.~\ref{sec:spin2}. One can see this because we apply a laser pulse to the state $\ket{0}^{\otimes N}$, which creates the state $\ket{0}\ket{1}\ket{0}^{\otimes N-2}$, and the two are coupled by the laser strength $\Omega_1(t)$. The Hamiltonian further couples this state to two states, $\ket{1}\ket{2}\ket{0}^{\otimes N-2}$ and $\ket{0}\ket{2}\ket{1}\ket{0}^{\otimes N-3}$; the latter then couples further, forming a chain of states. With a third laser on spin 1, coupling the states $\ket{1}$ and $\ket{2}$, the state $\ket{1}\ket{2}\ket{0}^{\otimes N-2}$ is coupled to $\ket{2}\ket{2}\ket{0}^{\otimes N-2}$ with the laser strength $\Omega_0(t)$. Consequently, the state transfer sequence requires $O(N+N^{1/3}\log\varepsilon)$, and thus a single gate, one- or two-qubit, can be achieved within a time $O(N^{7/3}+N^{5/3}\log\varepsilon)$, which is more efficient than the previous `fast' pulses, and derives all the benefits of robustness demonstrated in Sec.~\ref{sec:robust}.

\subsection{Cooling}

Naturally, it is desirable to be able to prepare the initial state of the system using the control available. With the described mapping between the UQI system $H_L+H_3$ and a state transfer chain $H_{TV}$, one would hope that the first step in preparing the initial state of the system would arise from application of the results in Sec.~\ref{sec:cooling}. The mapping would allow us to remove any of the $\ket{1}$ states from the $a$ system. Unfortunately, this is not the case -- we have only proved the mapping in the first excitation subspace. In order for the results to be applicable in higher excitation subspaces, one of two conditions needs to be fulfilled. Either the rotations $U$ need to commute with each other (i.e.~$[U_{i-1,i},U_{i,i+1}]=0$) or the state of the $b$ system would have to be in the maximally mixed state $\rho=\identity/2^N$. While this second condition may often be a good assumption, we note that a more generally applicable technique was described in \cite{Burgarth:07} where one simply measures one of the spins, and sets it to a fixed state (say $\ket{0}$), and repeating. Note that unlike the rest of \cite{Burgarth:07}, it is not necessary to use an ancillary system.

Once all the $\ket{1}$ excitations have been removed from the $a$ subsystem, the entire system is in an eigenstate of the Hamiltonian. However, we still have to prepare the $a$ system as $\ket{0}^{\otimes N}$ rather than an unknown superposition of $\ket{0}$ and $\ket{2}$ states. Consider the setting where all spins are $\ket{0}$ except for one, which is in the $\ket{2}$ state (this argument trivially generalizes to any arbitrary distribution of $\ket{2}$s). Now, each time we apply the standard state transfer protocol from spin 1 to spin $N$, the $\ket{2}$ shifts one spin towards spin 1 (and flips to a $\ket{0}$, where all other states have become $\ket{2}$). Thus, after no more than $N$ repetitions, the unwanted state has reached the control area and can be corrected. Once the $a$ spins are correctly initialized, we can implement the cycle operation on the $b$ qubits, setting each of them in turn to $\ket{0}$ as they pass through a control spin, ready to proceed with the computation.

\subsection{Fault Tolerance}

The architecture that we have described so far does not support fault-tolerant quantum computation. With only a fixed number of access points to the chain, the degree of parallelism is insufficient, which means that errors build up more quickly than they can be corrected. One can avoid this problem by allowing control of a regular array of spins \cite{UQI}. However, this is not the end of the problems. We will not dwell on these here, merely describe them, and possible solutions, for the sake of completeness, while noting that much of the structure of the scheme (limited parallelism, the need to correct both a classical and quantum region of the device, when the only way to interact with the quantum region is through the classical region) is very similar to global control schemes, in which fault tolerance is possible \cite{Kay:thesis}. Moreover, the present setting has the advantage that we can perform different actions (e.g.~error correction based on syndrome measurements) at each point of interaction with the chain, rather than needing the same interaction. As such, one might expect a vastly improved fault-tolerant threshold.

During the computation, no matter how well we protect the system, some noise will always enter. Typically, one assumes that these errors are single-spin rotations. Of course, if a single $b$ qubit is affected by noise, then standard error correction techniques are sufficient to correct the error. However, if an error were to flip, say, between the $\ket{0}$ and $\ket{2}$ states of an $a$ spin, this means that every time a global unitary is applied through the state transfer protocol, it is faulty on a specific pair of qubits. As with the cooling mechanism, this pair changes with each repetition, so that by the time the error has propagated to the control area, many of the computational qubits in the block will have been affected. The effects are somehow related, and it may be possible to design error correcting codes to compensate for this. In particular, our error detection sequence only needs to determine where the flip first occurred, which is very similar to detecting the edge of a pattern of spins, such as the boundary term in $\ket{11\ldots100\ldots0}$. This can be related via a local unitary transformation into detection of a single excitation \cite{kay-2006b}, which is precisely the task that standard error correcting codes are designed for. Alternatively, given that the error only persists for a finite time, then the affected region of the $b$ qubits is no larger than the separation between two control regions, and error correcting codes that are capable of correcting that many errors can be used (or the error correcting codes can be used on non-contiguous blocks of qubits). A similar effect will result if the initial Hamiltonian is subject to local perturbations, or if multiple $\ket{1}$ states were to appear in the $a$ region. Thus, it should be possible to handle all such issues, although it is not our priority in the present paper.

\section{Summary and Outlook}

We have presented a constructive technique that shows how simple one-dimensional Hamiltonians can be used for universal quantum computation, with control of only a small number of the spins. This construction is based on using state transfer in an auxiliary system and thus, when limited to a nearest-neighbor translationally invariant Hamiltonian necessarily requires a local Hilbert space dimension of at least 4 (2 for the auxiliary system, and 2 for the computational qubit). While control is over 3 spins of the system, each of the actions we require is restricted to acting on a two-dimensional subspace of an individual spin. Furthermore, we have shown how to incorporate a number of useful aspects such as robustness to control errors, a deterministic cooling protocol and the slow nature of the control fields.

Evidently, the intention of designing such a scheme is to make experimental implementation more feasible. However, more accurately, this work represents a limit of analytic constructive techniques, illustrating the abilities that are present in simple systems. The first experiments will require even simpler systems, such as a 1D Heisenberg chain of qubits, and, as such, one expects the control theory solutions to be the first to be experimentally tested, and the results here only become relevant when scaling becomes a critical issue in the design of control pulses in these simpler systems.

An interesting avenue for further theoretical work would be Hamiltonian simulation -- our constructions have been carefully tailored to give quantum gates that can be composed using the circuit model. However, one of the major applications of a quantum computer will be quantum simulation, where we attempt to simulate a variety of different Hamiltonians in a system which is easily controlled. Are there UQI systems where control over the spins influences an effective Hamiltonian evolution, without having to resort to a Trotter decomposition of the evolution? If resorting to a Trotter decomposition, one solution immediately suggests itself. Temporarily neglecting the `slow rotations' component, and taking the $a$ system to have 4 levels, a Hamiltonian
$$
H_{\text{sim}}=\sum_n\sum_{i=1}^3\ket{0i}\bra{i0}^a_{n,n+1}\otimes U_{\sigma_i}^b+h.c.,
$$
where
$
U_{\sigma}=e^{-i\delta t\sigma\otimes\sigma}
$
and $\sigma_i\in\{X,Y,Z\}$, allows simulation of a Hamiltonian $\sum_n(\lambda_xXX+\lambda_yYY+\lambda_zZZ)_{n,n+1}$ simply by selecting the correct ratio of the number of uses of the different excitations in the state transfer process. Furthermore, by applying the inverse (i.e.~by implementing state transfer on the $a$ system from spin $N$ to 1), we are able to generate composite pulse sequences to increase the accuracy of the simulation \cite{composite_pulses}.

{\em Acknowledgments:} This work is supported by DFG (FOR 635 and SFB 631), EU (SCALA), Clare College, Cambridge and the National Research Foundation \& Ministry of Education, Singapore.

\end{document}